\documentstyle[psfig,prb,aps]{revtex}
\begin{document} 
\draft

\twocolumn[\hsize\textwidth\columnwidth\hsize\csname
@twocolumnfalse\endcsname

\title{\bf Calculation of the Band Gap Energy and Study of Cross 
           Luminescence in Alkaline-Earth Dihalide Crystals.}
\author{Andr\'es AGUADO,\footnote{Email address: aguado@jmlopez.fam.cie.uva.es} 
Andr\'es AYUELA$^1$, Jos\'e M. L\'OPEZ, 
and Julio A. ALONSO}
\address{Departamento de F\'\i sica Te\'orica,
Universidad de Valladolid, Valladolid 47011, Spain}
\address{$^1$ Laboratory of Physics, Helsinki University of Technology,
02015 Espoo, Finland}
\maketitle

\begin{abstract}
The band gap energy as well as the possibility of cross luminescence processes
in alkaline-earth dihalide crystals have been calculated using the 
{\em ab initio} Perturbed-Ion (PI) model.
The gap is calculated in several ways: as a
difference between one-electron energy eigenvalues and as a
difference between total energies of appropriate electronic states of the
crystal, both at the HF level and with inclusion of Coulomb correlation effects.
In order to study the possibility of ocurrence of cross luminescence in these
materials, the energy difference
between the valence band and the outermost core band for some
representative crystals has been calculated.
Both calculated band gap energies and cross luminescence predictions compare
very well with the available experimental and theoretical results.

\end {abstract}
           
\vskip2pc]

\section{Introduction}

An important revival of the interest in luminescent materials
is presently observed.
This is mainly due to their practical applications in several fields,
like nuclear spectroscopy, dosymetry, or two-dimensional 
detectors used for example in medical
screens and crystallography,\cite{Bla95}$^)$ to mention a few.
Frequently these luminescent materials turn out to be single ionic crystals.
Alkaline-earth dihalide crystals (pure or doped) are between the most important 
inorganic scintillators.
Within this class of materials,
fluoride crystals with the fluorite structure ($CaF_2$, $SrF_2$ and $BaF_2$)
are the best known scintillators, and recent work has been
devoted to their study when doping with lanthanum \cite{Vis92}$^)$, $Mn^{2+}$
\cite{Pas95,Luc96}$^)$ and
$Ce^{3+}$ impurities.\cite{Vis93a,Vis93,Mei96,Riv97}$^)$
Chlorides
are less used materials, though recent theoretical work on the scintillation
properties of $SrCl_2:Ce^{3+}$ samples has been reported.\cite{Mer96}$^)$
Experimental studies of the luminescent emission from impurity centres in
$SrCl_2$ \cite{Tab92}$^)$ and $BaCl_2$ \cite{Tab89}$^)$ single crystals 
have also been reported.
Before
formulating a model for the description of doped complexes like $BaF_2:Ce^{3+}$,
fundamental information about the pure crystals should be compiled.
Here, as a first step towards a more profound comprehension of optically active
impurities, we have focused our interest on the description of the
pure materials.

The calculation of the energy band gap ($E_{gap}$) of these materials 
is important for several reasons:
when doping the crystals, it becomes necessary to locate the impurity levels
in the band gap of the pure crystal.
Furthermore, the gap is a very important quantity in the
first step of the scintillation process, namely absorption of
radiation leading to formation of electron-hole pairs. 
This step influences the global 
efficiency of the scintillator. Last but not the least, 
it is also important
from the theoretical point of view, and it gives information on the goodness of
the model.

The discovery of the fast luminescent
component in the emission spectrum of $BaF_2$ and the cross-luminescence
(CL) mechanism \cite{Kub88,Ito88}$^)$ has brought about a 
large activity in the field of scintillator
research.\cite{Eij94}$^)$ $BaF_2$, with a 0.8 ns CL component is
the inorganic scintillator with the fastest response. This makes $BaF_2$
a very attractive material for applications in positron emission tomography
(PET), where a good time resolution is of paramount importance in order to
supress random coincidences and to use time-of-flight information.\cite{Mos96}$^)$
$BaF_2$ has also been proposed as a candidate material for detectors
in high-energy physics \cite{Ale84,Val85}$^)$, most recently
at the new proton colliders SSC and LHC at CERN,\cite{Lec94}$^)$ where the important
requirement is to differenciate events from different bunch crossings. 
In the CL mechanism, an electron is first
promoted from the core band to the conduction band, leaving a hole
in the core band, and next a valence band electron recombines with that
hole giving rise to cross-luminescence and leaving a hole 
in the valence band (see Fig. 1). That hole can recombine then with the
electron promoted to the conduction band, leading to selftrapped exciton (STE)
emission.
An important parameter in the study of cross luminescence is the energy
separation between valence and core bands ($\Delta E_{VC}$), 
as it determines, for a given material,
whether the ocurrence of CL is possible or not.
Specifically, this energy difference must be smaller than the band gap 
energy\cite{Kub88,Ito88,Eij94}$^)$ (that is, $\Delta E_{CV} < E_{gap}$) in order to observe 
CL.
This is associated to the fact that, if the emission energy is smaller than
the band gap, photon reabsorption inside the crystal or Auger emission of
electrons are
not possible.

In the past, there has been a considerable deal of work devoted to 
calculate the band
structure of some of these ionic crystals, mainly those with
the fluorite structure.
\cite{Sta73,Jou76,Alb77,Rob79,Hea80,Kin81a,Kin81b,Eva89}$^)$
The purpose of the research carried out here is to
study pure
$AX_2$ crystals, where $A$ stands for $Mg$, $Ca$, $Sr$, $Ba$ and $X$ for
$F$, $Cl$, $Br$, $I$, by using instead a cluster approach. 
To this end we have used the
Perturbed Ion (PI) model.\cite{Lua90}$^)$
The present work is an extension of our own
previous research on ionic crystals, where
first we calculated the band gap energies of alkali halide crystals,\cite{Agu96}$^)$
and then studied several properties of the scintillators NaI:Tl$^+$ and
CsI:Tl$^+$.\cite{Agu98}$^)$
Given the great technological significance of cross luminescence processes, 
we have also
calculated the energy difference between the valence band and the
upmost core band
for some
$AX_2$ materials, and compared it with the relevant gap energy. This
allows to predict the occurrence of cross
luminescence in these solids.
Cluster approaches have been used in the past to study the CL mechanism in
alkaline-earth fluorides by
Andriessen {\em et al.} \cite{And91}$^)$, and also by Ikeda {\em et al.}
\cite{Ike97a,Ike97b}$^)$

The paper is structured as follows:
In the next section we show how to apply the PI model
to the calculation of gap energies between valence and conduction bands 
and between valence
and core bands. The results of the calculations are
presented and discussed in section III. Section IV presents our conclusions.

\section {Calculational method}

\subsection {Brief summary of the theoretical model}

A detailed description of the PI method has
been given in our previous
work on band gap energies of alkali halide crystals\cite{Agu96}$^)$ and in the
original works\cite{Lua90,Lua90b}$^)$, so here we just
give a brief resume.
In order to calculate the electronic structure of ionic solids, and
in particular the band gap energy, the traditional band theory approach
could be directly applied. For doped
crystals, however, the translational symmetry is lost, and the
Bloch's theorem does not apply.
Cluster models, which have been successfully
applied to the analysis of local
crystal properties, provide 
an alternative, although any cluster approach has some inherent difficulties
in dealing with delocalized conduction states
(see, for example, ref. \onlinecite{Bag94}).
The cluster approximation is based on
a partition of the solid into a ``cluster region'' 
and ``the rest of the crystal''.
The main issues one has to address when using this approximation are the
following : a) the election of the best partition 
of the crystal into ``cluster'' and
``host'' regions; b) a proper
quantum mechanical description of the finite cluster;
c) a precise description of the ions surrounding the cluster (this
is necessary for a proper account of cluster-lattice interactions); and d)
the consistency between the descriptions of the finite cluster and of the
surroundings.
As a matter of fact, issues c and d have received
less attention than issues a and b. Issue d is frequently neglected,
and the ``rest of the crystal'' is often simulated
just by using point charges.\cite{Bar83}$^)$ More accurate lattice models have
been considered,
for example by Winter, Pitzer and Temple,\cite{Win87a,Win87b}$^)$ who
introduced effective core potentials to describe the lattice cations nearest to
the cluster. The significance of the cluster-lattice consistency was 
stressed by Kunz and coworkers,\cite{Kun88}$^)$ who included the
so-called localizing potentials in the description of the cluster. Besides,
issue a is also a delicate point, as it is difficult to avoid surface effects
at the cluster boundary when studying bulk properties with cluster models.

In the PI model, the ``cluster'' is reduced to its minimum
size, a single ion, and cluster-lattice interactions are described in the
framework of the Theory of Electronic Separability (TES) of Huzinaga 
{\em et al.}\cite{Huz71,Huz73}$^)$
Being the ``cluster'' a single ion, boundary effects are avoided right from
the start.
Furthermore, the cluster approximation can be rigourously formulated within the
TES, as cluster-lattice orthogonality is a fundamental requirement of that
theory. Each ion in the lattice acts on the ``cluster'' density
through an effective potential which includes nuclear, Hartree and
exchange contributions. The electronic structure of each inequivalent ion
in the crystal is self-consistently determined,
thus avoiding the use of empirical parameters.\cite{Lua90}$^)$
Another feature of the PI model is that it does not invoke the
Molecular Orbital LCAO approximation.
The one-center character of the model
leads to a large computational-time saving compared to any multi-center
cluster approach.

The atomic-like orbitals used to describe each ion are expanded into a large 
set of Slater type (STO) basis functions\cite{Cle74a,Cle74b}$^)$ because 
of their superior performance.
The outputs of the PI model are a set of crystal-adapted wave functions for each
ion, which are fully consistent, in the Self-Consistent-Field
(SCF) sense, with the lattice potential, and the effective energies of the
corresponding ions in the field of the crystal lattice. The effective energy
is written as a sum of the net energies (or self-energies)
of the ions in the crystal and
the ion-lattice interaction energies. Once those quantities have been obtained,
the total crystal energy is obtained as a sum of monocentric terms, 
involving
net energies, and bicentric contributions coming from the interaction
between ions.\cite{Lua90}$^)$ A correlation energy correction following the
unrelaxed Coulomb-Hartree-Fock (uCHF) prescription\cite{Cle65,Cha89}$^)$ 
is added to the net
energies.\cite{Lua90b}$^)$
The PI method is particularly well suited to model an impurity center
in a crystal because we can
describe the ions surrounding the impurity with
lattice-consistent wave functions, rather than with Hartree-Fock free-ion
wave functions.
Applications of the PI model
to these problems can be consulted in refs. \onlinecite{Agu98,Lua92}.

\subsection {Band gaps}

The band gap energy can be defined as the difference between the two following
energies:\cite{Poo75a}$^)$ a) the energy necessary to ionize the crystal by
removing an electron from the top of the
valence band (the threshold energy $E_t$); b) the energy gained by putting
an electron at the bottom of the empty conduction band 
of the perfect crystal
(the electronic affinity
$\chi$):
\begin{equation}
E_{gap} = E_t - \chi.
\end{equation}
To calculate $\chi$ we have to model an
electron in the conduction band, that is in a delocalized
state. The strong ion-lattice
orthogonality required by the TES would force that electron to be localized on a
given ion.
We can, however, obtain a
reasonable
description of a delocalized conduction state by relaxing some of 
the orthogonality
requirements imposed by the TES (see below). The electronic affinity
$\chi$ is always a small quantity, close to zero for all these ionic
materials,\cite{Poo75a,Poo75b,Pon78,McC85}$^)$
and our description gives values of the correct order of magnitude
(a few tenths of eV).
As typical errors in measured gaps are $\sim$ 0.5 eV,
the quantitative determination of the band gap energy is not critically affected
by our approximations. All the calculations are performed at the experimental
crystal structures.\cite{Wyc63,Bus70,Bra63,Bec81}$^)$

We proceed now to calculate the
band gap energy in several ways. In a first calculation
we approximate the threshold energy $E_t$ by the energy eigenvalue (with
opposite sign) of the highest occupied anionic orbital
obtained in a PI calculation for the pure crystal at the HF level. This
corresponds to the Koopmans' approximation to the ionization potential.
To obtain $\chi$, we simulate a singly charged alkaline-earth cation $A^+$
as an impurity in the field created by the pure crystal ($A^+:A^{2+}X^-_2$),
and equate $\chi$ to the eigenvalue (with opposite sign)
of the outermost s-orbital of that cation. To allow for the
delocalization of the orbital over a substantial region of the crystal we
enlarge the basis set taken from Clementi and Roetti\cite{Cle74a,Cle74b}$^)$ for $A^+$
with some diffuse functions and relax the orthogonality requirement 
between that s orbital
and the surrounding lattice.
This eigenvalue is close to zero (in no case is $-\epsilon$
greater than 0.2 eV).
According to the definition (eq. 1), our first calculation of the gap reads:
\begin{equation}
E_{gap} = \epsilon_{ns}(A^+:A^{2+}X^-_2) - \epsilon_{mp}(X^-:A^{2+}X^-_2)
\end{equation}
where n=3,4,5,6 (m=2,3,4,5) for $Mg$, $Ca$, $Sr$ and $Ba$ ($F$, $Cl$, $Br$ and
$I$), respectively. In recent studies, de Boer and de Groot \cite{deB98a,deB98b}$^)$
have shown that the conduction band of those ionic crystals formed with alkali
or alkaline-earth cations is associated mainly to anionic levels. Thus, we have
checked the effect of centering the conduction electron on an anionic site.
As long as we enlarge the anion basis set with enough diffuse functions, the
value of $\chi$ is almost insensitive to the election of specific center for the
conduction electron.

In a second calculation, 
the band gap energy is obtained as a difference of total energies, as in a 
typical $\Delta$SCF calculation. This approach
includes orbital relaxation effects, improving thus 
over Koopmans' approximation.
The independent processes of removing an electron from a
halogen anion and of placing that electron on the conduction band read now: 
\begin{equation}
E_t = E_{crystal}(X^0:A^{2+}X^-_2) - E_{crystal}(A^{2+}X^-_2),
\end{equation}
and
\begin{equation}
\chi = E_{crystal}(A^{2+}X^-_2) - E_{crystal}(A^+:A^{2+}X^-_2),
\end{equation}
respectively. 
$E_{crystal}(X^0:A^{2+}X^-_2)$ represents the energy of the crystal
with a single neutral halogen 
impurity and $E_{crystal}(A^+:A^{2+}X^-_2)$ the energy of
a crystal with a delocalized electron although centered on an 
alkaline-earth ion (see above). 
In the process of calculating the energies of the systems $X^0:A^{2+}X^-_2$ and
$A^+:A^{2+}X^-_2$,
the ions surrounding the
impurities are described with the lattice-consistent wave functions obtained in
the PI calculation of the pure crystal.
With these assumptions, $E_t$ and $\chi$ reduce to a
difference of effective energies
\begin{equation}
E_t = E_{eff}(X^0:A^{2+}X^-_2) - E_{eff}(X^-:A^{2+}X^-_2) 
\end{equation}
\begin{equation}
\chi = E_{eff}(A^{2+}:A^{2+}X^-_2) - E_{eff}(A^+:A^{2+}X^-_2).
\end{equation}
$E_{eff}(X^0:A^{2+}X^-_2)$ is the effective energy of a neutral halogen atom in
an otherwise perfect crystal $A^{2+}X^-_2$, and a similar interpretation holds
for the other effective energies (for details, consult refs.
\onlinecite{Agu96} and \onlinecite{Map92}).
$E_t$ gives the main contribution to
$E_{gap}$, while $\chi$ only provides a
small correction of magnitude $\sim 0.2$ eV, 
as in the previous calculation.
The $\Delta$SCF calculations are performed at the HF and uCHF
levels; the 
second ones include, besides orbital relaxation effects,
correlation corrections to the HF energy. 

\subsection{Cross luminescence}

We extend now the previous expressions to the calculation of 
$\Delta E_{VC}$, the energy difference
between valence and core bands. In some
$AX_2$ crystals (see section III.C below), the 
core band nearest to the valence band is associated
to the outermost p-orbitals of the $A^{2+}$ cation, but in some others it is 
associated to the
outermost s-orbitals of the $X^-$ anion (when the 
anionic s-eigenvalue is less negative than the cationic p-eigenvalue). Our first
calculation is again a simple difference between energy eigenvalues
(Koopmans' approximation).
For the first
case:
\begin{equation} 
\Delta E_{VC} = \epsilon_{mp}(X^-) - \epsilon_{n-1,p}(A^{2+}),
\end{equation}
and in the second case:
\begin{equation}
\Delta E_{VC} = \epsilon_{mp}(X^-) - \epsilon_{ms}(X^-),
\end{equation}
with a notation consistent with
eq.(2).
We can improve the calculation
of $\Delta E_{VC}$ allowing for the relaxation of the hole states
created in the curse of the CL process (see Fig. 1). 
We obtain first the core binding
energy, $E_C$. Allowing for orbital relaxation, the expression for $E_C$ is
similar to eq.(5):
\begin{equation}
E_C = E_{eff}(A^{3+}:A^{2+}X^-_2) - E_{eff}(A^{2+}:A^{2+}X^-_2),
\end{equation}
if the core band is formed by the outermost cationic p-orbitals, or
\begin{equation}
E_C = E_{eff}(X^0(ms^1mp^6):A^{2+}X^-_2) - E_{eff}(X^-:A^{2+}X^-_2),
\end{equation}
if the core band is formed by the outermost anionic s-orbitals
(the notation $X^0(ms^1mp^6)$ indicates the electronic configuration of the
halogen species after removing one electron from the ms orbital).
$E_C$ is the energy necessary to remove an electron from
the core band out of the crystal. 
By substracting from this energy the threshold energy of
the crystal, given in eq.(5),
we obtain the following approximation for the energy difference
between core and valence bands:
\begin{equation}
\Delta E_{VC} = E_C - E_t.
\end{equation}
As before, the $\Delta$SCF calculations are carried out at the HF and uCHF
levels of theory.

\section{Results and discussion}

An important part of the whole process involves to solve the electronic
structure of
the pure crystal. The PI model is a general method for dealing with
crystalline compounds of any spatial group, and the only inputs required 
are the spatial group and the lattice constants.
$CaF_2$, $SrF_2$, $BaF_2$ and $SrCl_2$
adopt the fluorite structure ($Fm3m$ in the international notation for
spatial groups), $MgF_2$  a rutile-type structure ($P4_2/mnm$),
$CaCl_2$ and $CaBr_2$ adopt distorted rutile structures ($Pnnm$), 
$MgCl_2$ has the
cadmium chloride structure ($R\overline{3}m$), and $MgBr_2$, $MgI_2$ and $CaI_2$
the cadmium iodide structure ($P\overline{3}m1$). Finally $BaCl_2$, $SrBr_2$, $BaBr_2$
and $BaI_2$ adopt the $PbCl_2$ structure ($Pbnm$).\cite{Wyc63}$^)$
The treatment of complicated crystalline structures is not a challenge for the
PI calculations.
Experimental geometries have been obtained from ref.\onlinecite{Wyc63}, except those
of $BaCl_2$, $BaBr_2$ and $BaI_2$, which have been taken from 
refs.\onlinecite{Bus70},\onlinecite{Bra63} and \onlinecite{Bec81}, respectively. 
In the $Pbnm$ structures
the two anions are in slightly nonequivalent positions, so they have different
eigenvalues and effective energies. To calculate the threshold energy
in these cases we will take out from the crystal the
least bound p-electron, which corresponds to an anion $X^-$ in a definite
site. In all the other cases, anions are in equivalent positions.

\subsection{Band gaps}

Results of the calculated band gaps of fifteen alkaline-earth dihalide crystals
are presented 
in Table I. We give the calculated gaps at three diferent levels of
theory and compare these with experimental results whenever they are available.
We have collected the experimental results from several sources.
\cite{Eij94,Poo75a,Poo75b,Tho73,Sug74}$^)$
The $\Delta \epsilon_{HF}$ gaps from eq. (2)
show the expected behavior in ionic solids,
namely, the band gap decreases when moving down the periodic table along the
alkaline-earth (halogen) column, leaving fixed the halogen (alkaline-earth)
ion. The influence of a different spatial symmetry does not affect at all
these trends, so it seems that its effect on band gap energies is small.
Comparison to experimental values shows a systematic overestimation of the gap. 
The gaps obtained by 
the $\Delta$SCF(HF) procedure are smaller than those calculated by
substracting eigenvalues. Relaxation of the halogen atom in response to
the removal of one 
electron from the anion (which is included in $E_t$ at this level
but not in the previous calculation) is the main reason responsible for
this effect. The influence of orbital relaxation on $\chi$ is rather 
negligible.
We 
observe for the $\Delta$SCF(HF) method an underestimation of the
band gap energies of fluorides but still an overestimation for chlorides. 
Inclusion of Coulomb correlation effects is crucial to achieve
quantitatively good results, as the $\Delta$SCF(uCHF) calculations
show. Upon inclusion of
correlation, the band gap energies change in the correct direction both for
fluorides and chlorides, increasing the gap in the first group and
decreasing it in the second group. The larger response of the 
anionic density cloud to correlation effects is
also responsible for this improvement. On the other hand, $\chi$ remains
practically unchanged, as before.

Theoretical studies of $CaF_2$ and $MgF_2$ 
have been
carried out by Catti {\em et al.}\cite{Cat91a,Cat91b}$^)$
using CRYSTAL, an
{\em ab initio} periodic Hartree-Fock program.\cite{Dov89} $^)$
They have obtained
ground state properties, like equilibrium geometries or elastic constants, and
also have shown the calculated 
band structure for these crystals. The band
gap energies obtained are 21 eV and 20 eV for $CaF_2$ and $MgF_2$, respectively.
The HF approximation tends to overestimate the gap in ionic
crystals.\cite{Kun82}$^)$ At first sight, it seems that
these results should be compared to our 
$\Delta \epsilon_{HF}$ calculation, which, although still 
overestimating the gap,
gives closer agreement with experiment.
However, both calculations should not be directly compared. 
In refs. \onlinecite{Cat91a,Cat91b} the gap is identified with the energy difference
between the highest occupied and the lowest unnoccupied orbitals, that is a
$LUMO - HOMO$ calculation. In our $\Delta \epsilon_{HF}$ calculation, while
$E_t$ can be identified still
with the $HOMO$, $\chi$ can not be identified with the
$LUMO$ anymore, as it is the eigenvalue (with opposite sign) 
of an $occupied$ orbital.
We think that this fact,
together with the appropriate inclusion of diffuse 
basis functions in the description of the
conduction state to simulate delocalization (diffuse valence orbitals on 
cations were not 
included in the above mentioned works) are the main
reasons for our improvement over their results. This
shows that even an energy difference between eigenvalues can be a quite good
approximation when the conduction state is more realistically described. 

More recently, Ikeda {\em et al.}\cite{Ike97a}$^)$ have calculated the band gaps of
CaF$_2$, SrF$_2$ and Ba$F_2$ by using active clusters of different sizes
embedded in a field of point charges representing the Coulomb potential
created by the crystals. The electronic structure of those active clusters is
solved by the use of the discrete variational (DV) X$\alpha$ method.\cite{Ada78}$^)$
The band gaps obtained by Ikeda {\em et al.} show some dependence on the
size of the active cluster employed, but for a fixed cluster size, show the
correct trend, that is a systematic decrease of the band gap when changing the
cation size from Ca$^{2+}$ to Ba$^{2+}$. A meaningful 
quantitative comparison with our set 
of results is not direct, because their calculation identifies the gap with a
difference between one-particle electronic levels (like our 
$\Delta\epsilon_{HF}$ calculation), but including correlation effects
at the same time footing as exchange through the X$\alpha$ model. Nevertheless,
the gaps they obtain with the (F$_7$A$_4$)$^+$ cluster are almost identical to
our most accurate $\Delta$SCF(uCHF) results.

\subsection{Cross luminescence}

Now we turn to the calculation of the energy differences involved in cross
luminescence processes. We show in Table II
the energy eigenvalues (with
opposite sign) of the
core orbitals nearest to the valence band, for four representative crystals.
We note that the outermost core band is formed by cationic alkaline-earth p-states
in $SrF_2$, $BaF_2$ and $BaCl_2$, and by
anionic halide s-states in $SrCl_2$. So the energy of the
core band level
($E_C$) is, in the Koopmans' approximation, identified with the 3s
eigenvalue of $Cl^-$ in $SrCl_2$, with the 5p eigenvalue of $Ba^{2+}$ in
$BaF_2$ and $BaCl_2$ and with the 4p eigenvalue of $Sr^{2+}$ in $SrF_2$.
The energy differences $\Delta E_{VC}$ are shown at three levels
of theory in Table III, together with experimental results 
for fluorides as given in
ref. \onlinecite{Eij94}.
The $\Delta \epsilon_{HF}$ calculation underestimates
$\Delta E_{VC}$ in these materials, because the
Koopmans' approximation overestimates the threshold energy $E_t$ more than
$E_C$.
We can appreciate this item more
clearly when passing from the 
$\Delta \epsilon_{HF}$ to the $\Delta$SCF(HF) calculation.
Here, contrary to the band gap calculation (see Table I), 
$\Delta E_{CV}$ increases in $SrF_2$, $BaF_2$ and $BaCl_2$, 
showing that anionic orbital relaxation effects are
more important than cationic ones: both $E_t$ and $E_C$ are lower than in the 
$\Delta \epsilon_{HF}$ calculation, but the lowering of $E_t$ is more
important, resulting in an increasing of $\Delta E_{CV}$. In
$SrCl_2$, such behaviour is not observed, and this is due to the fact that
the core band is an anionic band. The $\Delta$SCF(HF) result
lowers slightly the magnitude of $\Delta E_{VC}$ for this crystal. This effect is very
small, however, showing that the error in applying
Koopmans' approximation is nearly the same for $E_t$ and $E_C$ in this case. 
The $\Delta$SCF(HF) results include orbital relaxation,
and the different relaxations of core and valence holes are
responsible for all the effects discussed. 
Finally, the effect of
Coulomb correlation ($\Delta$SCF(uCHF) column) is, except for $BaCl_2$,
very small, showing that the corrections introduced in the calculation of
$E_v$ and $E_C$ are similar, and tend to cancel out in the energy difference.
Then it seems that correlation tends to shift occupied bands by a rather constant
amount with respect to the conduction band, leaving energy differences
between occupied bands unaffected, at least for fluorides.

In the fifth column of Table III we show the possibility of occurrence of CL
in these crystals. 
As we explained in the introduction,
if the energy difference $\Delta E_{VC}$
between the valence and the outermost core band is
smaller than the energy gap $E_{gap}$ between conduction and valence bands, 
then it is possible to
observe CL.
This condition holds in $BaF_2$, so we can assert that CL is possible, as it
has been experimentally observed.\cite{Ito88,Eij94}$^)$
Regarding $SrF_2$, we obtain that 
$\Delta E_{VC}$ is larger than $E_{gap}$ by 1.7 eV, so cross luminescence is
not predicted by our calculations.
The situation is even more unfavourable for
$SrCl_2$ and $BaCl_2$ crystals (for which we obtain a
value for $\Delta E_{VC}$ larger than the corresponding band gap
energy by 8 eV) as well as for the rest of
$AX_2$ crystals not shown in table III.
The combination of fluorine
and barium, which gives the most ionic compound between those studied in
this work, gives a favourable limiting case
for CL applications. Barium is the cation with the smallest ionization
potential in the alkaline-earth series, so it is easier to remove an electron
from the 5p orbital of $Ba^{2+}$ than from other alkaline-earth cations. 
This leads to a lower value of $E_C$ for barium compounds. On the other hand, 
fluorides show the largest band gap energies between alkaline-earth 
dihalides systems (as in simple alkali halide crystals \cite{Agu96}$^)$),
because it is more difficult to remove an electron from a $F^-$ anion than from
any other halide anion. 
As a consequence, the difference in eq.(11) is the lowest in
magnitude.

The cluster model calculations performed by Ikeda {\em et al.} \cite{Ike97a}$^)$
lead to the same conclusions, although the specific value obtained for
$\Delta$E$_{VC}$ again depends a little on the size of the active cluster.
They obtain a band gap for BaF$_2$ which is 3 eV larger than $\Delta$E$_{VC}$,
so their calculations predict also the occurrence of cross luminescence for
BaF$_2$. 
The band gap of CaF$_2$ is smaller than
the energy difference between the valence and core bands.
For SrF$_2$, the values of E$_{gap}$ and $\Delta$E$_{VC}$ are similar,
again in agreement with our results. 

\section{Conclusions}

We have calculated the band gap energies of fifteen alkaline-earth dihalide
$AX_2$ crystals. For some of these crystals we have also calculated the
energy difference between core and valence bands in order to study
the possibility of cross luminescence (CL) processes. We have followed two
different ways to estimate these quantities: first, we have taken differences
between energy eigenvalues, and next we have carried out $\Delta$SCF
calculations, both at the HF level and with inclusion of correlation effects.
To this end we have used the Perturbed Ion (PI) model,
supplemented with some additional assumptions to deal with an electron
at the bottom
of the conduction band.
This is a cluster-type model which achieves full lattice-cluster 
consistency.
Correlation effects have been included using
a model proposed by Clementi.\cite{Cha89}$^)$
Within the PI model we can
deal with crystals of any spatial point group symmetry without any problem.
The 
significance of an accurate estimate of the band gap energy in these crystals
as a preliminary requirement for the study of luminescence properties has been
pointed out, as well as the importance of the energy difference between core and
valence bands in the CL process. The calculation of the last quantity is
possible because our calculation is an all-electron calculation and does not
invoke any approximations like frozen-core treatments.
Orbital relaxation of the core hole is also allowed.
However, one has to recognize that our method gives no dispersion for the
core or valence bands so the applicability is restricted to cases when these
bands are narrow enough.

Overall we have obtained a rather good estimation of the energy band gaps,
with a computational effort which is much less than that required by any
other cluster multi-center approach or by standard band structure methods.
Gaps compare quantitatively well with experimental results whenever they are
available, and favourably to previous theoretical calculations for
fluorides.
The CL mechanism has been predicted as possible only
in $BaF_2$, in agreement with the experimental observations.

$\;$

$\;$

{\bf Acknowledgments}:
This work has been supported by DGES(Grant PB95-0720-C02-01) and Junta de
Castilla y Le\'on (Grant JCL-VA70/99).
A. Aguado is greatful to JCL
for a postgraduate fellowship.

\pagebreak

{\bf Captions of figures}

$\;$

$\;$

{\bf Figure 1}

Schematic view of the cross luminescence process.

\onecolumn[\hsize\textwidth\columnwidth\hsize\csname
@onecolumnfalse\endcsname

\begin{figure}
\psfig{figure=crosslum.epsi,height=0.8\linewidth,width=\linewidth,angle=270}
\end{figure}

\pagebreak

\begin {table}
\begin {center}
\caption{Calculated band gaps,
compared to available experimental
values. SPG gives the spatial point group of the crystal, in the
international notation. $\Delta \epsilon_{HF}$ is a difference between
one-particle orbital energy eigenvalues. $\Delta$SCF refers to a difference
between the total energies of the 
crystal in appropriate electronic states (see
text).
All energies are given in eV.}

\begin {tabular} {|c|c|c|c|c|c|c|}
\hline
& $Crystal$ & SPG & $\Delta \epsilon_{HF}$ & $\Delta$SCF(HF) & $\Delta$SCF(uCHF) & Exp. \\ 
\hline
& $MgF_2$ & $P4_2/mnm$ & 14.3 & 10.8 & 12.8 & 12.4$^{\em a}$ \\
& $CaF_2$ & $Fm3m$ & 12.7 & 10.0 & 11.9 & 12.0$^{\em b}$  \\ 
& $SrF_2$ & $Fm3m$ & 12.4 & 9.7 & 11.5 & 11.1$^{\em b}$  \\ 
& $BaF_2$ & $Fm3m$ & 12.0 & 9.2 & 11.0 & 10.5$^{\em b}$  \\ 
& $MgCl_2$ & $R\overline3m$ & 10.7 & 9.2 & 7.9 & 7.5$^{\em c}$  \\
& $CaCl_2$ & $Pnnm$ & 10.2 & 9.1 & 8.0 & 6.9$^{\em c}$  \\
& $SrCl_2$ & $Fm3m$ & 9.6 & 8.5 & 7.4 & 7.5$^{\em c}$  \\
& $BaCl_2$ & $Pbnm$ & 8.9 & 7.8 & 6.8 & 7.0$^{\em c}$  \\
& $MgBr_2$ & $P\overline3m1$ & 10.5 & 9.3 & 7.6 & - \\
& $CaBr_2$ & $Pnnm$ & 9.7 & 9.0 & 6.7 & - \\
& $SrBr_2$ & $Pbnm$ & 8.5 & 8.0 & 6.5 & - \\
& $BaBr_2$ & $Pnma$ & 8.3 & 7.7 & 6.0 & - \\
& $MgI_2$ & $P\overline3m1$ & 9.9 & 9.2 & 7.2 & - \\
& $CaI_2$ & $P\overline3m1$ & 9.2 & 8.4 & 6.3 & - \\
& $BaI_2$ & $Pbnm$ & 8.1 & 7.4 & 5.4 & -

\end {tabular}
\end {center}

$^{\em a}$: Ref. \onlinecite{Tho73}

$^{\em b}$: Ref. \onlinecite{Poo75a}

$^{\em c}$: Ref. \onlinecite{Sug74} 

\end {table}

\begin {table}
\begin {center}
\caption{One-particle orbital energy eigenvalues for the outermost p-orbital of the
$A^{2+}$ cation and the outermost s-orbital of the $X^-$ anion.
All energies are given in eV.}

\begin {tabular} {|c|c|c|c|}
\hline
 & $Crystal$ & $\epsilon(A^{2+})$ & $\epsilon(X^-)$ \\ 
\hline
& $SrF_2$ & 23.6 (4p) & 36.9 (2s) \\ 
& $BaF_2$ & 17.7 (5p) & 36.3 (2s) \\ 
& $SrCl_2$ & 27.0 (4p) & 25.6 (3s) \\
& $BaCl_2$ & 20.4 (5p) & 24.9 (3s) 

\end {tabular}
\end {center}

\end {table}

\begin {table}
\begin {center}
\caption{Energy differences between the upmost core band level and the
valence band level calculated from different theoretical methods (as in Table I), 
and possibility of cross luminescence (CL). 
All energies in eV.}

\begin {tabular} {|c|c|c|c|c|c|c|}
\hline
 & $Crystal$ & $\Delta \epsilon_{HF}$ & $\Delta$SCF(HF) & $\Delta$SCF(uCHF) & CL & Exp. \cite{Eij94}$^)$ \\ 
\hline
& $SrF_2$ & 11.2 & 13.2 & 13.2 & no & ? \\ 
& $BaF_2$ & 5.7 & 7.8 & 7.7 & yes  & yes \\ 
& $SrCl_2$ & 16.0 & 15.8 & 15.4 & no  & - \\
& $BaCl_2$ & 11.4 & 12.2 & 14.8 & no  & -

\end {tabular}
\end {center}

\end {table}

\end{document}